\documentclass[proceedings, preprint]{rmaa}

\usepackage{paralist}

\usepackage{psfrag,color}

\SetYear{2008}
\SetConfTitle{Magnetic Fields in the Universe II}

\title{Magnetic Fields and Star Formation in Spiral Galaxies}

\author{ Marita Krause,\altaffilmark{1} }

\altaffiltext{1}{Max-Planck-Institute for Radioastronomie, Auf dem
H\"ugel 69, 53121 Bonn, Germany (mkrause@mpifr-bonn.mpg.de).}

\shortauthor{Marita Krause}
\shorttitle{Magnetic fields in galaxies}

\listofauthors{Marita Krause}
\indexauthor{Krause, M.}

\abstract{The main observational results from radio continuum and polarization observations about the magnetic field strength and large-scale pattern for face-on and edge-on spiral galaxies are summarized and compared within our sample of galaxies of different morphological types, inclinations, and star formation rates (SFR). We found that galaxies with low SFR have higher thermal fractions/ smaller synchrotron fractions than those with normal or high SFR. Adopting an equipartition model, we conclude that the nonthermal radio emission and the \emph{total magnetic field} strength grow nonlinearly with SFR, while the \emph{regular} magnetic field \emph{strength} does not seem to depend on SFR.\

We also studied the magnetic field structure and disk thicknesses in highly inclined (edge-on) galaxies. We found in four galaxies that -despite their different radio appearance- the vertical scale heights for both, the thin and thick disk/halo, are about equal (0.3/1.8~kpc at 4.75~GHz), independently of their different SFR. This implies that all these galaxies host a galactic wind, in which the bulk velocity of the cosmic rays (CR) is determined by the total field strength within the galactic disk. The galaxies in our sample also show a similar large-scale magnetic field configuration, parallel to the midplane and X-shaped further away from the disk plane, independent of Hubble type and SFR in the disk. Hence we conclude that also the \emph{large-scale} magnetic field \emph{pattern} does not depend on the amount of SFR.}

\resumen{}

\addkeyword{galaxies: magnetic fields}
\addkeyword{galaxies: spiral}
\addkeyword{galaxies: individual (NGC~891)}
\addkeyword{galaxies: individual (NGC~4631)}
\addkeyword{galaxies: halos}
\addkeyword{polarization}

\begin{document}
\maketitle

\section{Introduction}
\label{sec:intro}
Radio observations of the continuum emission in the cm-wavelength
regime are the best way to study magnetic fields in galaxies. The total
intensity of the synchrotron emission gives the strength of the total magnetic field. The linearly polarized intensity reveals the strength and the structure of the resolved regular field in the sky plane (i.e. perpendicular to
the line of sight).  However, the observed polarization vectors suffer
Faraday rotation and depolarization on their way from the radiation's
origin to us. Correction for Faraday rotation is possible with
observations at two -- or better more -- wavelengths by determining the
rotation measure RM (being proportional to $\int n_{\rm e} B_{\parallel}
dl$). $B_{\parallel}$ is the coherent magnetic field parallel to the
line of sight, and its sign gives the direction of this magnetic field
component. Both field components, parallel and perpendicular to the
line of sight, enable us in principle to deduce a 3-dimensional
picture of the large-scale magnetic field.\
Note, however, that the polarized intensity is only sensitive to the
field orientation, i.e. it does not distinguish between parallel and antiparallel field directions in the plane of the sky, whereas the RM is large for parallel fields along the line of sight, but zero for parallel and antiparallel fields (of equal strength).\
Magnetic fields consist of regular and turbulent components. The total
magnetic field strength in a galaxy can be estimated from the
nonthermal radio emission under the assumption of equipartition
between the energies of the magnetic field and the relativistic
particles (the so-called {\em energy equipartition}) as described in
Beck \& Krause (2005).\

\section{Observational Results of Magnetic Fields}

\subsection{Spiral Galaxies seen face-on}
\label{sec:face-on}

The mean equipartition value for the total magnetic field strength for
a sample of 74 spiral galaxies observed by Niklas (1995) is on average
$9\,\pm 3\,\mu$G. It can, however, reach locally higher values {\em
within} the spiral arms of up to $20\,\mu$G.  Strongly interacting
galaxies or galaxies with a high star formation rate tend to have
generally stronger total magnetic fields.
The strength of the regular magnetic fields in spiral galaxies
(observed with a spatial resolution of a few 100pc) are typically
1--5~$\mu$G, and may reach locally values up to $10 - 15\,\mu$G as e.g. in NGC6946 (Beck 2007) and M51 (Fletcher et al. 2008).\

The turbulent magnetic field is typically strongest along the optical
spiral arms, whereas the regular fields are strongest in between the
optical spiral arms, or at the inner edge of the density-wave spiral
arm. Sometimes, the interarmed region is filled smoothly with regular
fields, in other cases the large-scale field form long filaments of polarized intensity like in IC342 (Krause et al. 1989, Krause 1993)
or so-called {\em magnetic spiral arms} like in NGC6946 (Beck \& Hoernes 1996).\

The magnetic lines of the large-scale field form generally a spiral
pattern with pitch-angles from $10\arcdeg$ to $40\arcdeg$ which are
similar to the pitch angles of the optical spiral arms. Further,
spiral magnetic fields have even been observed in flocculent and
irregular galaxies.\

\subsection{Spiral Galaxies seen edge-on}
\label{sec:edge-on}
\begin{figure}[htb]
\centering
\includegraphics[bb = 26 27 457 609,width=7.0cm,clip=]{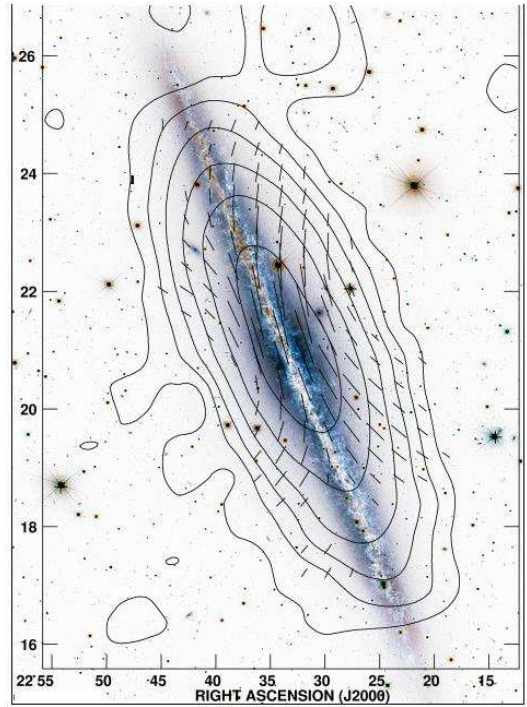}
\caption{Radio continuum emission of the edge-on spiral galaxy NGC~891 at
$\lambda3.6$cm (8.35~GHz) with the 100-m Effelsberg telescope with a
resolution of $84 \arcsec~HPBW$. The contours give the total intensities,
the vectors the intrinsic magnetic field orientation (Copyright: MPIfR Bonn).
The radio map is overlayed on an optical  image of NGC~891 from the
Canada-France-Hawaii Telescope / (c)1999 CFHT/Coelum.
}
\label{n891}
\end{figure}

Several edge-on galaxies of different Hubble type and covering a wide range
in SFR were observed with high sensitivity in radio continuum and linear polarization.
These observations show that the magnetic field structure is mainly {\em parallel
to the disk} along the midplane of the disk (with the only exception of NGC4631)
as expected from observations of face-on galaxies and their magnetic field
amplification by the action of a mean-field $\alpha\Omega$-dynamo (e.g. Beck
et al. 1996). Away from the disk the magnetic field has also vertical components
increasing with distance from the disk and with radius (Krause 2004, Soida 2005,
Krause et al. 2006, Heesen et al. 2009b). Hence, the large-scale magnetic field
looks X-shaped away from the plane. As an example we show the magnetic field
vectors of the edge-on galaxie NGC~891 together with the radio continuum emission
at $\lambda3.6$cm (made with the Effelsberg 100-m telescope \footnote{The
Effelsberg 100-m telescope is operated by the ailsshMax-Planck-Institut f\"{u}r
Radioastronomie (MPIfR) in Bonn.} overlayed on an optical image of the galaxy
in Fig. \ref{n891}.

The rotation measures (RM) determined between our $\lambda3.6$cm-map and the $\lambda6.2$cm-map of Sukumar \& Allen (1991) vary smoothly along the major axis of the galaxy with positive values in the southern part and negative in the northern part of the galaxy. This indicates a large-scale toroidal field within the galactic disk. As the RMs have equal signs above and below the disk along the major axis, the azimuthal magnetic field in NGC~891 seems to be an {\em even} axisymmetric spiral field (ASS) (Krause 1990) as generated by a mean-field $\alpha \Omega-$~dynamo (cf. Sect.~\ref{sec:regular field}).\

From the HI- and CO-line observations (Rupen 1991, Garc\'{\i}a-Burillo \& Gu\'{e}lin 1995) we know that the rotation velocities in NGC~891 are also negative in the northern part of the galaxy and positive in the southern one. Assuming trailing optical spiral arms (and a magnetic field {\em orientation} primarily parallel to the opical arms as observed in face-on galaxies) we can infer that the {\em direction} of the large-scale ASS field in NGC~891 is {\em outwards}. Together with NGC~4254 (Chy{\.z}y 2008) and NGC~5775 (Soida et al. in prep.) these are the first examples of outwards directed ASS fields. The other four galaxies for which the direction of the ASS could be determined up to now (M31, IC342, NGC~253, and NGC~6949) were all inwards directed, which could not be explained up to now (F. Krause \& Beck 1998). The three counterexamples NGC~891, NGC~4254, and NGC~5775 balances the statistics.\

\section{Total Magnetic Field Strength and Star Formation}
\label{sec:total field}

Observations of a sample of three late-type galaxies with low surface-brightness and the radio-weak edge-on galaxy NGC~5907 (all with a low SFR) revealed that they all have an unusually high thermal fraction and weak total and regular magnetic fields (Chy{\.z}y 2007, Dumke et al. 2000). However, these objects still follow the total radio-FIR correlation, extending it to the lowest values measured so far. Hence, these galaxies have a lower fraction of synchrotron emission than galaxies with higher SFR. It is already known that the thermal intensity is proportional to the SFR. Our findings fits to the equipartition model for the radio-FIR correlation (Niklas \& Beck 1997), according to which the nonthermal emission increases
$\propto SFR^{\approx 1.4}$ and the \emph{total} magnetic field strength as well as the ratio of nonthermal-to-thermal emission increases with $\propto SFR^{\approx 0.4}$.\

\begin{figure}[htb]
\centering
\includegraphics[bb = 26 28 471 466,width=7.5cm,clip=]{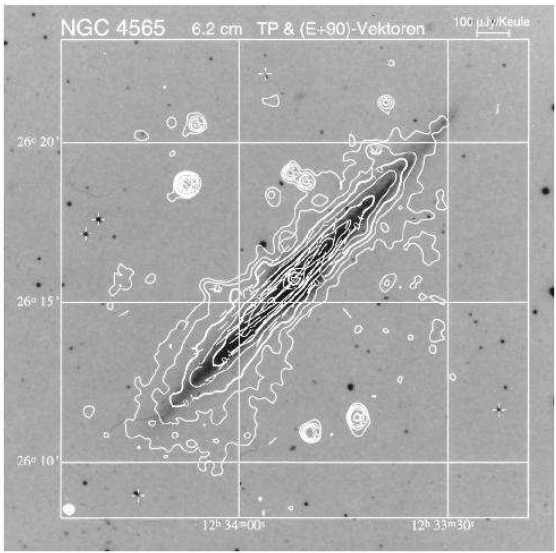}
\caption{Radio continuum emission of the edge-on spiral galaxy NGC~4565 at $\lambda6.2$cm (4.85~GHz), combined from observations with the VLA and the 100-m Effelsberg telescope with $20 \arcsec~HPBW $ (Dumke 1997). The contours give the total intensities, the vectors the intrinsic magnetic field orientation. The radio map is overlayed on an optical image of NGC~4565. (Copyright: MPIfR Bonn)
}
\label{n4565}
\end{figure}

We also determined the exponential scale heights of the total power emission at $\lambda6$~cm for four edge-on galaxies (NGC~253, NGC~891, NGC~3628, NGC~4565) for which we have combined interferometer and single-dish data (VLA and the 100-m Effelsberg) at this wavelength. The map of NGC~4565 is shown in Fig. \ref{n4565}. In spite of their different intensities and extents of the radio emission, the {\em scale heights} of the thin disk and the thick disk/halo are similar in this sample (300pc and 1.8kpc) (Dumke \& Krause 1998, Heesen et al. 2009a). We stress that our sample includes the brightest halo observed so far, NGC~253, with a very high SFR as well as one of the weakest halos, NGC~4565, with a small SFR.\

In the case of NGC~253, Heesen et al. (2009a) found that the local scaleheights of the radio emission are linear functions of the cosmic ray electron (CRE) synchrotron lifetimes in the underlying disk. This implies that the cosmic ray transport is mainly in vertical direction with a cosmic ray bulk speed (determined as the ratio of the total power scaleheight and the CRE synchrotron lifetimes) of at least 280~km/s. The cosmic ray bulk speed can be best described as a convective transport of cosmic rays (CR) in a galactic wind (as proposed e.g. by Breitschwerdt et al. 1991), and the above deduced correlations give observational support for the existence of such a galactic wind.\

The fact that we observe similar averaged scaleheights at $\lambda6$~cm for the four galaxies mentioned above imply --together with the assumption that the synchroton lifetimes ($\propto \rm B ^{-2}$) determine the disk thicknesses--, imply that the CR bulk speed (or galactic wind velocity) is proportional to $\rm B ^2$, and hence proportionel to $\rm {SFR} ^{0.8}$.

\section{Regular Magnetic Field and Star Formation}
\label{sec:regular field}

The polarized intensity in spiral galaxies allows to estimate the regular magnetic field strength and the observed degree of polarization P (as determined by the ratio of polarized intensity and total intensity) is a measure of the degree of uniformity of the magnetic field. As discussed in Sect.~\ref{sec:total field} the low brightness edge-on galaxy NGC~5907 has a weak total and polarized intensity, a high thermal fraction and a low SFR. On the other hand, NGC~891 (see Fig. \ref{n891}) has strong total and polarized intensities, a normal thermal fraction and strong star formation. Surprisingly, the degree of polarization at $\lambda6$~cm at a resolution of $84 \arcsec~HPBW $ (half power beam width) is equal in both galaxies, $\rm P = 3.2 \%$. When averaged the Stokes I-, U-, and Q-maps over the total galaxy (as described in Stil et al. 2008), the global degree of polarization in NGC~5907 is $\rm P = 2.2 \%$, whereas in NGC~891 it is only $\rm P = 1.6 \%$.\

Similarily, we integrated the polarization properties in 41 nearby spiral galaxies and found that (independently of inclination effects) the degree of polarization is lower ($ < 4\%$) for more luminous galaxies, in particular those for $ L_{4.8} > 2 \times 10^{21}~\rm{W Hz^{-1}}$ (Stil et al. 2009). The brightest galaxies are those with the highest SFR. Of course, the mean-field dynamo needs star formation and supernova remnants as the driving force for velocities in vertical direction. From our observations, however, we conclude that stronger star formation seems to reduce the magnetic field regularity.\

On kpc-scales, Chy{\.z}y (2008) analyzed the correlation between magnetic field regularity and SFR locally within one galaxy, NGC~4254. While he found a strong correlation between the total field strength and the local SFR, he found an anticorrelation of magnetic field regularity with SFR and could not detect any correlation between the regular field strength and the local SFR.\

\begin{figure}[htb]
\centering
\includegraphics[bb = 26 28 465 380,width=7.5cm,clip=]{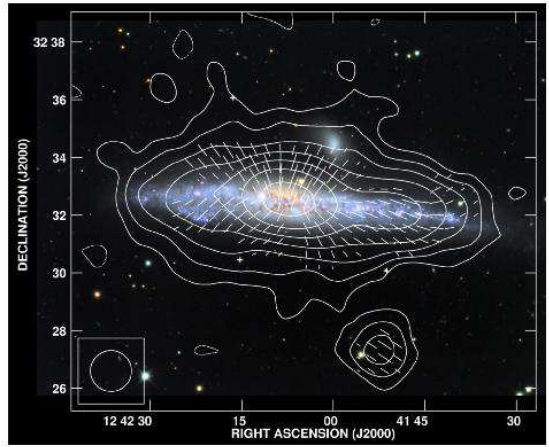}
\caption{Radio continuum emission of the edge-on spiral galaxy NGC~4631 at
$\lambda3.6$cm (8.35~GHz) with the 100-m Effelsberg telescope with $84 \arcsec~HPBW $. The contours in give the total intensities, the vectors the intrinsic magnetic field orientation. The radio map is overlayed on an optical image of NGC~4631 taken with the Misti Mountain Observatory. (Copyright: MPIfR Bonn)
}
\label{n4631}
\end{figure}

In our sample of 11 observed edge-on galaxies we found in all of them with one exception (in the inner part of NGC~4631) mainly a disk-parallel magnetic field together with the X-shaped poloidal field in the halo. Our sample include spiral galaxies of very different Hubble type and SFR, ranging from $0.5 \le \rm{SFR} \le 27$. The disk-parallel magnetic field along the galactic disk is the expected edge-on projection of the spiral-field within the disk as observed in face-on galaxies. It is generally thought to be generated by a mean-field $\alpha \Omega-$~dynamo for which the most easily excited field pattern is the axismmetric spiral (ASS) field (e.g. Ruzmaikin et al. 1988). The dynamo acts most effectively in regions of strong differential rotation in the disk. NGC~4631, however, which has a disk-parallel magnetic field along the disk only at radii $\ge$ 5kpc, shows for smaller radii a vertical large-scale field also in the midplane of the disk, as visible in Fig. \ref{n4631}.
The inner $\approx 5$~kpc ($140 \arcsec$) is just the region of NGC~4631 where the rotation curve rises nearly rigidly (Golla \& Wielebinski 1994), hence the mean-field $\alpha \Omega-$~dynamo may not work effectively in this inner part of NGC~4631 and hence may not amplify a disk-parallel large-scale field.\

The even ASS magnetic disk field (as discussed in Sect.~\ref{sec:edge-on}) is -- according to the mean-field dynamo theory -- accompanied with a quadrupolar poloidal field, which is, however, by a factor of about 10 weaker than the toroidal disk field. This means that the poloidal part of the ASS dynamo mode alone cannot explain the observed X-shaped structures in edge-on galaxies as the field strengths there seems to be comparable to the large-scale disk field strengths. Model calculations of the mean-field $\alpha\Omega$-dynamo for a disk surrounded by a spherical halo including a {\em galactic wind} (Brandenburg et al. 1993) simulated similar field configurations as the observed ones. Such a galactic wind (which has already been observationally deduced e.g. as described in Sect.~\ref{sec:total field}) can also solve the helicity problem of dynamo action (e.g. Sur et al. 2007). Recent numerical dynamo simulations indicate a self-generation of vertical gas velocities (D. Elstner, private communication).\

As a summary, we conclude that -- though  a high SFR in the disk increases the total magnetic field in the disk and halo -- it cannot change the global field configuration nor influences the global scale heights of the radio emission.  The similar scale heights and magnetic field configurations in these galaxies imply that the total magnetic field strengths regulate the galactic wind velocities.\\

\end{document}